\documentclass[12pt,draftclsnofoot, onecolumn]{IEEEtran}
\usepackage{bbm}
\usepackage{amsfonts,longtable}
\usepackage[normalem]{ulem}
\usepackage{mathrsfs}
\usepackage{graphicx,cite,epsfig,amssymb,amsmath,color}

\newcommand{\tabincell}[2]{\begin{tabular}{@{}#1@{}}#2\end{tabular}}

\begin{document}

\title{Index Modulation for 5G:\\ Striving to Do More with Less}
\author{\normalsize
Xiang Cheng,~\IEEEmembership{Senior Member,~IEEE,} Meng Zhang, Miaowen Wen,~\IEEEmembership{Member,~IEEE,}\\
and Liuqing Yang,~\IEEEmembership{Fellow,~IEEE}\\
\vspace{1cm}
\thanks{ This work was supported in part by the National Natural Science Foundation of China under grant 61622101, 61571020, and 61501190,
the Ministry National Key Research and Development Project under Grant 2016YFE0123100, the open research fund of the State Key Laboratory of Integrated Services Networks under Grant ISN18-14, and the National Science Foundation under grant number CNS-1343189 and DMS-1521746.}
\thanks{Xiang Cheng is with State Key Laboratory of Advanced Optical Communication Systems and Networks, School of Electronics and Computer Science, Peking University, Beijing 100871, China and State Key Laboratory of Integrated Services Networks, Xidian University, Xi'an 710000, China. (Email: xiangcheng@pku.edu.cn).}
\thanks{Meng Zhang is with China Telecom Technology Innovation Center, Beijing 102209, China (Email: zhangmeng.bri@chinatelecom.cn).}
\thanks{Miaowen Wen is with School of Electronic and Information Engineering, South China University of Technology, Guangzhou 510640, China (Email: eemwwen@scut.edu.cn).}
\thanks{Liuqing Yang is with Department of Electrical and Computer Engineering, Colorado State University, Fort Collins, CO 80523, USA (Email: lqyang@engr.colostate.edu).}
\vspace{+1cm}
\begin{abstract}
The fifth generation (5G) wireless communications brag both high spectrum efficiency and high energy efficiency. To meet the requirements, various new techniques have been proposed. Among these, the recently-emerging index modulation has attracted significant interests. By judiciously activating a subset of certain communication {building blocks, such as} antenna, subcarrier and time slot, index modulation is claimed to have the potential to meet the challenging 5G needs. In this article, we will discuss index modulation and its general and specific representations, enhancements, and potential applications in various 5G scenarios. The objective is to reveal whether, and how, index modulation may strive for more performance gains with less medium resource occupation.
\end{abstract}
\vspace{+0.2cm}
}
\date{\today}
\renewcommand{\baselinestretch}{1.2}
\thispagestyle{empty} \maketitle \thispagestyle{empty}

\newpage

\IEEEpeerreviewmaketitle

\section{Introduction}
The explosive increase of mobile data services and the use of smart phones require the 5G network to support higher spectrum efficiency (SE), higher energy efficiency (EE), and higher mobility. The currently employed modulation techniques based on multi-input-multi-output orthogonal frequency division multiplexing (MIMO-OFDM), unfortunately, are not sufficient in satisfying such 5G requirements. {Conventional MIMO} may achieve high SE with massive antennas, but has compromised EE due to the scaled power consumption of a large number of RF chains. The OFDM modulation is prone to {Doppler-induced} intercarrier interference (ICI). Its inherent high peak-to-average power ratio (PAPR)  {also necessitates} expensive power amplifiers. Novel advanced modulation techniques are therefore very much needed. To this end, the recently emerging index modulation (IM) techniques arise as a promising candidate that has the potential to meet the 5G requirements{\cite{B16}}.

IM refers to a family of modulation techniques that rely on the activation states of some {resources/building blocks} for information embedding. The { resources/building blocks} can be either physical, e.g., antenna, subcarrier, time slot, and frequency carrier, or virtual, e.g., virtual parallel channels, signal constellation, space-time matrix, and antenna activation order. A distinct feature of IM is that part of the information is implicitly embedded into the  transmitted signal. Consequently, a much smaller number of RF chains is required for a massive MIMO configuration via the random selection of transmit antennas according to the information bits. This  { so-termed} spatial modulated (SM-)MIMO \cite{MHASL14} has been proved to strike a favorable tradeoff between SE and EE. One may also deactivate some subcarriers and use the activation states of subcarriers to convey additional information.  The resultant index modulated (IM-)OFDM technique \cite{MXMBH16} improves the SE and EE while alleviating  PAPR  and facilitating  easier ICI mitigation thanks to  empty subcarriers.

{ The basic concept and recent development of SM-MIMO and IM-OFDM were presented in \cite{B16}. Different from the approach therein, we aim to establish a unifying framework for general IM techniques, and provide indepth discussions on the SE and EE issues.} Note that both SM-MIMO and  IM-OFDM end up using {\bf{less}} medium resources relative to traditional MIMO-OFDM, in that only a subset of transmit antennas or frequency subcarriers are activated. Nevertheless, they tend to achieve {\bf{more}} EE and SE that are urged by 5G. It is then a curious question that whether IM can naturally do more with less. To address this intriguing question, in this article, we make an endeavor to investigate the general notion of IM which subsumes SM-MIMO and IM-OFDM as representative special cases. We systematically establish a general framework unifying space, time, and frequency , in which IM can be applied individually, combinatorially, and jointly. We then put existing forms of IM into this general framework in order to facilitate systematic delineation of their  pros and cons. From these, we aspire to shed light on whether IM can {\bf really do more with less.}

We then move on to discuss  possibilities of enhancing IM performance via, e.g., in-phase/quadrature IM, precoded IM, and diversity-enhancing IM. The objective is to reveal how these techniques can ultimately help IM to { further improve the SE and EE.}

All discussions in this article will be intimately coupled with core 5G application scenarios involving massive MIMO, full duplex, cooperative communications, and high mobility. Together with a unifying perspective for existing works in this field, future directions that can potentially fill blanks in this general framework will also be presented.

\section{The IM Family}
\begin{figure}[p]
  \centering
  \includegraphics[width=6in]{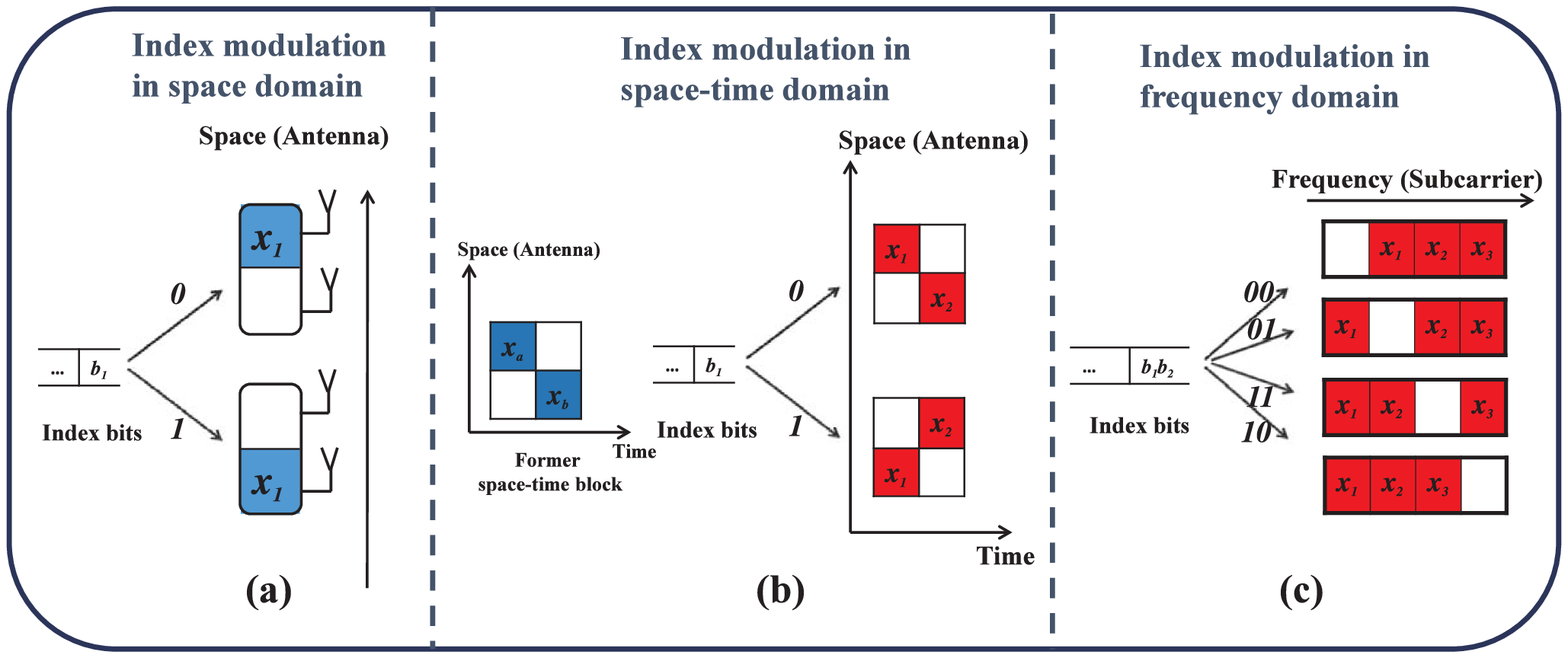}\\
  \caption{Index mapping procedures of: a) Spatial Modulation; b) Differential Spatial Modulation; c) IM-OFDM (4 subcarriers per group with 1 inactive one), where colored elements are set to be active according to the index bits $b_i$ and carry constellation symbols $x_i$, and the blank elements remain idle in the transmission. The positions of the active elements with the same color are determined by the same group of index bits.}\label{sysa}
\end{figure}

Existing IM schemes are mainly carried out in  space, time, and frequency slots, or the combination of them, among which the most marked similarity lies in the utilization of the active radio resources to carry information. {Generally speaking, IM splits the information bits into  index bits and  constellation bits. The  {former} determine which portion of the radio resources (antennas, subcarriers etc.)  {is} active, and the {latter} are mapped to conventional constellation symbols that are to be carried by the active resources. Suppose that $k$ out of the  $n$ resource indices are  active. The system SE in terms of bps/Hz  {is}:
\begin{align}\label{eq1}
{S_{IM}} = \mathop {\underline {\frac{1}{n}{{\log }_2}\left( {\mathbb{C}_n^k} \right)} }\limits_{{\text{index bits}}}  + \mathop {\underline {\frac{k}{n}{{\log }_2}\left( M \right)} }\limits_{{\text{constellation bits}}}
\end{align}
where ${\mathbb{C}_n^k}$ denotes the binomial coefficient, and $M$ is the cardinality of the symbol constellation.} Fig. \ref{sysa} demonstrates the index mapping  {of} existing IM schemes in different domains, which will be discussed  {next}.
\subsection{IM in  {Space}}
SM is a representative  IM technique in the space domain. As illustrated in Fig. \ref{sysa} (a), it works with a single RF chain and {conveys information via} the antenna index  \cite{MHASL14}. The {SM} signal constellation  consists of : 1) the conventional quadrature amplitude modulation constellation; and 2) the antenna index constellation.  Conveying information via the active antenna index, SM enjoys spatial multiplexing gain with a single-RF chain transmitter. {Hence},  SM is more energy efficient and  {its} detection complexity  {is lower than} vertical Bell Labs layered space-time (V-BLAST) systems. One drawback of SM is that the information  conveyed by the antenna index increases  {in} $\log_2 N_t$, where $N_t$ is the transmit antenna number. When $N_t$ is  large, SM suffers from compromised SE.  
The idea of SM can also  {accommodate} multiple active transmit antennas to improve  SE,  {giving rise to} generalized SM (G-SM). However, it requires multiple RF chains and induces inter-channel interference.

\subsection{IM in the {Space-Time}}
Transmitting signals across multiple time slots  {facilitate} transmit diversity {in MIMO systems}. Previous focus was on  space-time matrix {designs} that  strike the largest coding and diversity gains  with satisfactory receiver complexity. Recently, interests shift towards the exploration of the space-time resource to convey information. Differential SM (D-SM), an IM technique in the space-time domain, is a representative example \cite{YXMLHB13}. The index mapping  of D-SM is shown in Fig. \ref{sysa} (b){, which determines the current antenna activation order according to that in the former space-time block and the current index bits}. As  a differential solution to SM,  {D-SM} avoids the challenging channel estimation inherent to SM. Research  revealed that D-SM results in {no more than}  3dB performance loss compared with SM  \cite{YXMLHB13}.

{Like} SM, D-SM operates with a single RF chain,  {but bypasses} the accurate estimation of the channel state information (CSI)  {that is computationally demanding}.


\subsection{IM in {Frequency}}
IM-OFDM is a representative  IM technique in the frequency domain, which extends the  SM {principle} to OFDM subcarriers \cite{MXMBH16}.  In IM-OFDM, \emph{not} all subcarriers  {carry} information symbols, and the indices of inactive subcarriers  {convey} information  via IM , as illustrated in Fig. \ref{sysa} (c).

To date,  {several} work have demonstrated that IM-OFDM can outperform plain OFDM in terms of the bit error rate (BER) under the same SE. It is revealed in \cite{MXMBH16} that: 1) IM-OFDM achieves the maximum rate if and only if the subcarriers within each group experience independent fading {via e.g.,} interleaved grouping; 2) IM-OFDM with interleaved grouping can achieve an up to 3 dB signal-to-noise (SNR) gain over plain OFDM for small $M$ (typically $2,4$), though this improvement becomes smaller and even diminishes as $M$ grows; 3) The advantage of IM-OFDM over plain OFDM can be maximized by choosing a specific number of inactive subcarriers {(typically 1, or 2)},  and is more noticeable under PSK input.


\subsection{IM in {Space-Frequency}}
MIMO-OFDM has already been widely accepted as one fundamental technique in current wireless communication systems. To incorporate IM into MIMO-OFDM, one straightforward way is to carry out conventional modulation in one domain (space or frequency/subcarrier), and  superimpose IM in the other domain. For example, using all subcarriers, while implementing independent SM  {per} subcarrier leads to SM-OFDM; whereas using all antennas but implementing independent subcarrier activation  {per each} antenna would result in MIMO-OFDM with IM (MIMO-OFDM-IM)\cite{c4}. However, these architectures entail high detection complexity.  Low-complexity decoder design remains an open  problem.

\begin{figure}[p]
  \centering
  \includegraphics[width=6in]{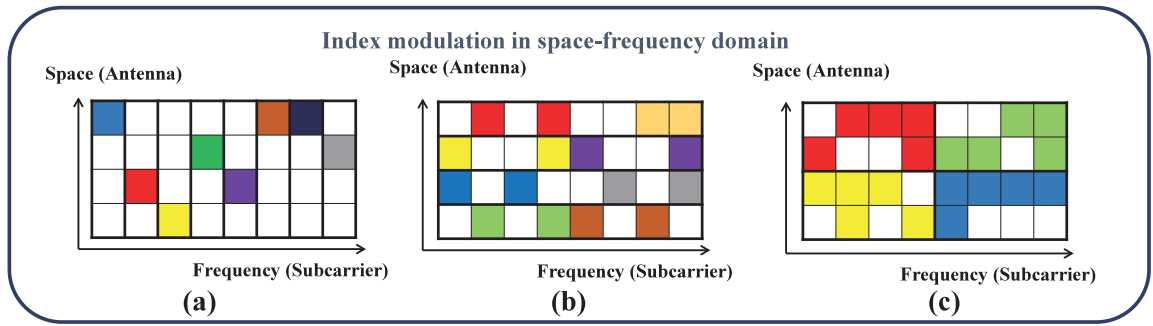}\\
  \caption{Index mapping strategies of: a)Spatial Modulation OFDM; b) MIMO-OFDM with IM; c) Generalized Space-Frequency IM, where colored elements are set to be active and carry constellation symbols and the blank elements remain idle in the transmission. The positions of the active elements with the same color are determined by the same group of index bits.}\label{sysaa}
\end{figure}

Recently, generalized space-frequency IM (GSFIM)\cite{c5} was proposed.  Instead of treating space and frequency domains independently, the active elements are jointly selected.  In GSFIM, a group of transmit antennas are  {activated} by the information bits and the active space-frequency elements  are then jointly selected according to the remaining information bits. Therefore, it  {is} the generalization of SM-OFDM and MIMO-OFDM-IM.  The index mapping  of SM-OFDM, MIMO-OFDM-IM, and GSFIM are shown in Fig. \ref{sysaa} (a)-(c).

\begin{figure}[p]
  \centering
  \includegraphics[width=6in]{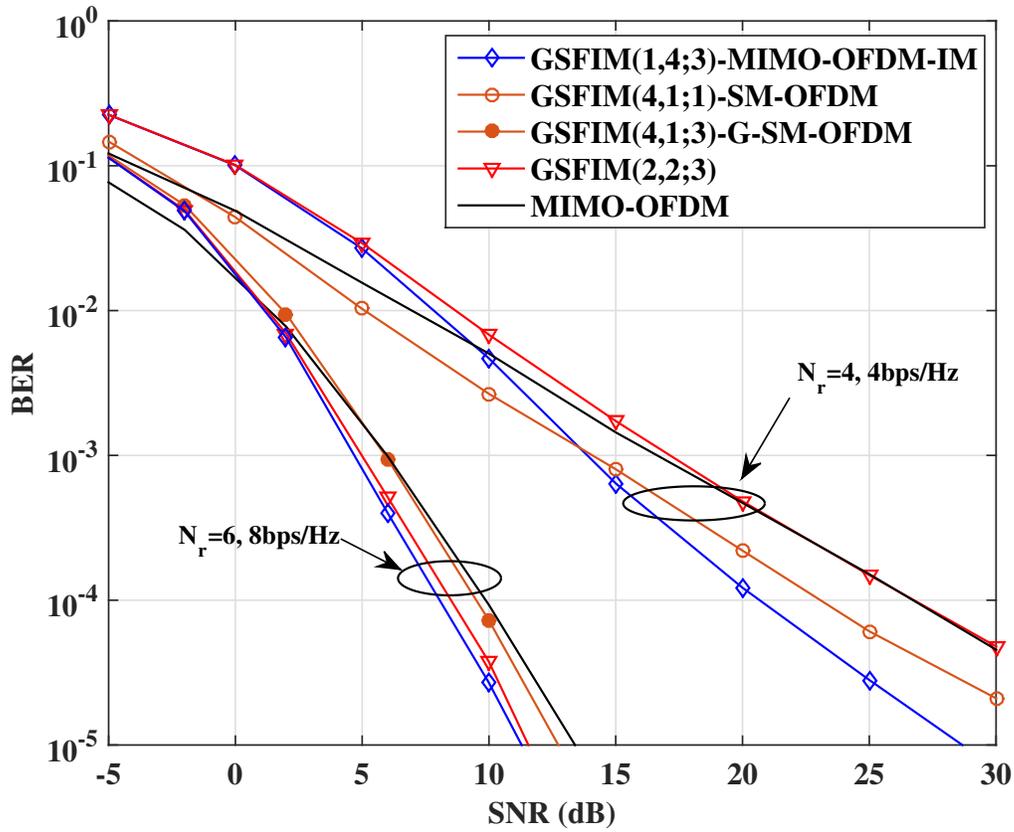}\\
  {\caption{Bit error ratio performance of MIMO-OFDM-IM {(BPSK, QPSK)}, GSFIM-OFDM {(BPSK, QPSK)}, and SM-OFDM {(QPSK, 64PSK)} with 4 transmit antennas and MMSE detector at 4 bps/Hz and 8 bps/Hz. {The parameter GSFIM($M,N;k$) means a GSFIM system with $k$ active elements in an $M$-antenna $\times$ $N$-subcarrier space-frequency group.  The curves of conventional V-BLAST MIMO-OFDM {(BPSK, QPSK)} are also given as a benchmark. Perfect CSI at the receiver.}}\label{sim1}}
\end{figure}

To fully exploit the benefit of IM in MIMO-OFDM,  comparison of  existing schemes is {important}. However, to date,  {such} study is still missing.  In Fig. \ref{sim1},  we { show} the bit error rate (BER) performance of MIMO-OFDM-IM, GSFIM, and SM-OFDM with 4 transmit antennas and practical minimum mean square error (MMSE) detector at  4 bps/Hz and 8 bps/Hz {over i.i.d. frequency selective Rayleigh  channels,  with perfect CSI}. Since MIMO-OFDM-IM and SM-OFDM are special cases of GSFIM,  all antennas  {are} active in GSFIM, so that  IM in space, frequency or the combination of them {can all be tested}. {At low data rate}, SM-OFDM has the best performance at low SNR and MIMO-OFDM-IM performs the best at high SNR. {Although all IM schemes perform worse at low SNR  due to the erroneous detection of the indices, they} demonstrate better error performance than V-BLAST MIMO-OFDM at medium-to-high SNR.  {At} 8 bps/Hz, SM-OFDM becomes the worst . However, when the frequency selectivity is not  significant, MIMO-OFDM-IM and GSFIM might suffer from performance degradation. {Hence,} SM-OFDM is more robust at lower data rate and modest frequency selectivity, MIMO-OFDM-IM and GSFIM are more favorable at higher data rates and more severe frequency selectivity.

\begin{figure}[p]
  \centering
  \includegraphics[width=4.8in]{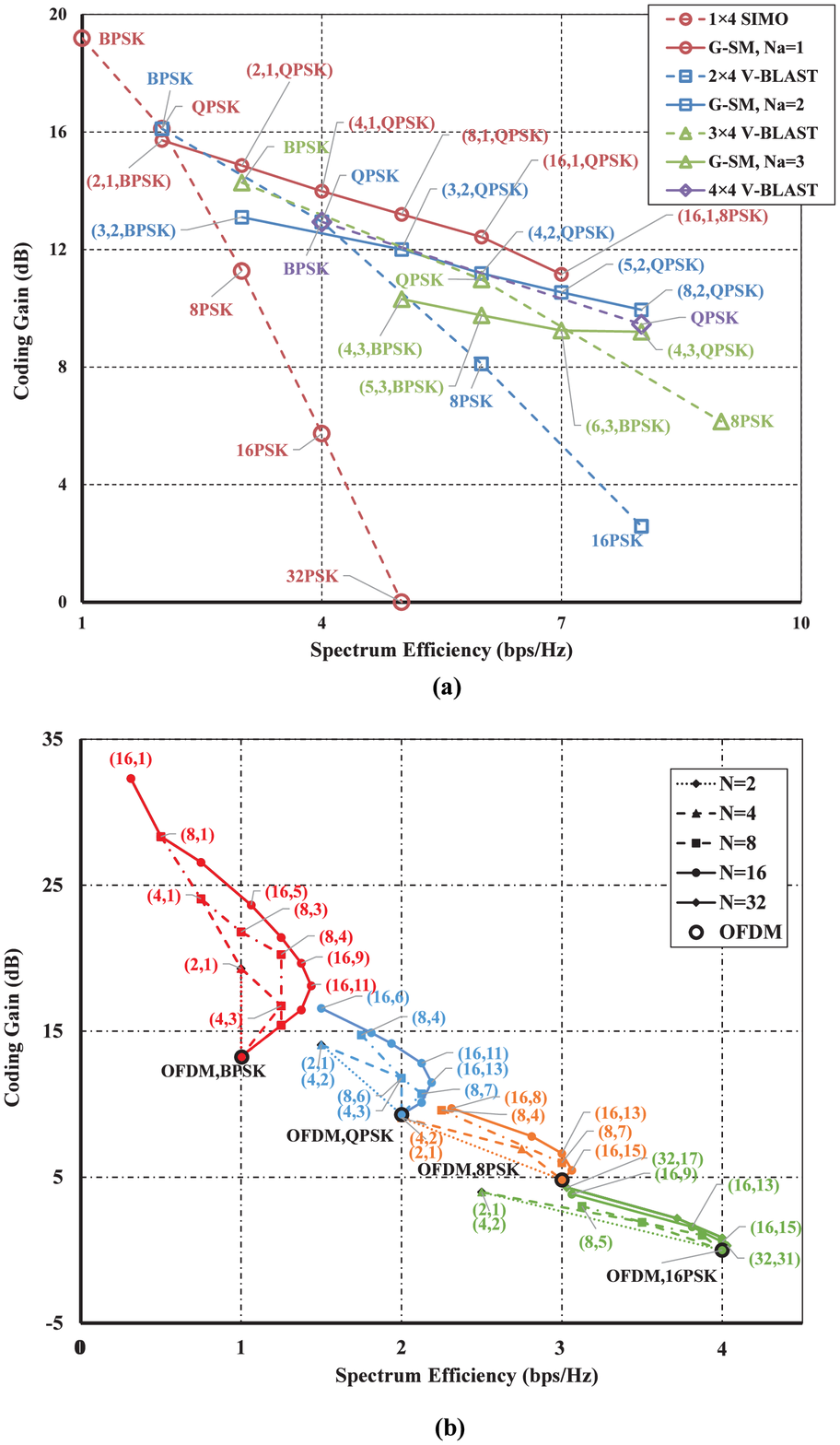}\\
  {\caption{Coding gain vs. spectrum efficiency for: a) V-BLAST and G-SM with 4 receive antennas and $1, 2, 3$ transmitter RF chains; b) OFDM and IM-OFDM{, where each curve of IM-OFDM shares the same modulation type with the point for OFDM it is connected with}.}\label{smseee}}
\end{figure}

IM activates a subset of antennas and/or subcarriers and uses the subset indices and/or activation order as part of the signal modulation. In this sense, IM is engaging less medium resource, in contrast with conventional schemes when all antennas and  subcarriers are activated all the time. Nevertheless, striving for improved EE and SE with less medium engagement, IM awaits not only optimized design in space-frequency domains, but also thorough investigation in the exploitation of time-domain.
{\section{Evaluation of the SE and EE}
In this section, we investigate the SE and EE issues in IM systems. To evaluate the EE, we derive the coding gain of the IM schemes{, $G_c$, which can be obtained by evaluating the high SNR approximation of the theoretical average bit error probability,
\begin{equation}\label{eq2}
{\text{ABEP}}\mathop  \propto \limits_{\gamma  \to \infty } {\left( {{G_c}\gamma } \right)^{ - {G_d}}}
\end{equation}
where $\gamma$ denotes the SNR, and $G_d$ is the coding gain.} Note that in practice, the system energy consumption should also  {account for} computational complexity and circuit/component power efficiency. Since the transmitter side RF chains contribute the largest portion to the system power consumption, if the schemes have similar hardware costs, e.g., the same number of RF chains, and also similar computational complexity, {the EE is solely determined by the coding gain, which determines the uncoded error performance, and indicates the target RF transmit power to guarantee reliable transmission.}
\subsection{SE and EE in MIMO with IM}
In MIMO systems, the conventional V-BLAST  is considered as the benchmark because it shares the same diversity order and similar decoding/encoding complexity with G-SM with the same number of RF chains  and maximum likelihood (ML) detection.

Fig. \ref{smseee} (a) shows the coding gain vs.  SE achieved by V-BLAST and G-SM with 4 receive antennas and PSK  in i.i.d. Rayleigh  environments. The coding gain is calculated with respect to the $1\times 4$ SIMO system with 32PSK modulation using the union bound. The G-SM system with $N_t$ transmit antennas, $N_a$ transmitter RF chains and $M$-PSK  is referred to as $(N_t, N_a, M{\text {PSK}})$. Three scenarios with $N_a=1,2,3$ are considered.
\begin{itemize}
  \item With fixed number of transmitter RF chains, G-SM can flexibly adjust the system SE by varying both the transmit antenna number and the modulation order, while V-BLAST can only change the modulation order.
  \item By increasing transmit antenna number without  {increasing} RF chains, G-SM benefits from moderate coding gain degradation, {e.g.,} higher EE, as SE increases.
  \item G-SM reduces the required RF chain number while achieving similar coding gain and SE.
\end{itemize}

\subsection{SE and EE in OFDM with IM}
Similar to G-SM, in IM-OFDM systems, the  SE improvement over OFDM can be attributed to the additional index bits mapped to the subcarrier activation patterns. However, the increase of EE is obtained differently. This is because IM in frequency cannot reduce the required RF chain number, and the EE gain solely {results} from coding gain.

In Fig. \ref{smseee} (b), the coding gain of OFDM/IM-OFDM with respect to OFDM with 16PSK is plotted as a function of the achievable SE. The IM-OFDM system with $A$ active subcarriers out of $N$ subcarriers per group is denoted by $(N,A)$. The channel is assumed to be frequency selective Rayleigh.  The conclusions are summarized  {below}.
\begin{itemize}
  \item The coding gain provided by IM is proportional to the ratio of inactive subcarriers and index bits.
  \item For a fixed modulation order, IM improves the SE and the coding gain simultaneously in comparison with  {plain} OFDM. But both gains degrade as the modulation order increases due to the decrease of the share of index bits among all information bits.
  \item {When considering the realistic frequency selectivity of the channel, for approximate SE and coding gain, smaller group size is more preferable to realize the theoretical performance}.
\end{itemize}

 {From all these, we can see that} IM in both space and frequency domains can provide EE and SE gains that are primarily facilitated by the index bits. In conclusion, from the perspective of achievable SE and RF transmit power, IM indeed achieves higher SE and EE while using less RF chains and activating less space and frequency resources.
}
\section{Enhancements of IM}
\begin{figure}[p]
  \centering
  \includegraphics[width=6in]{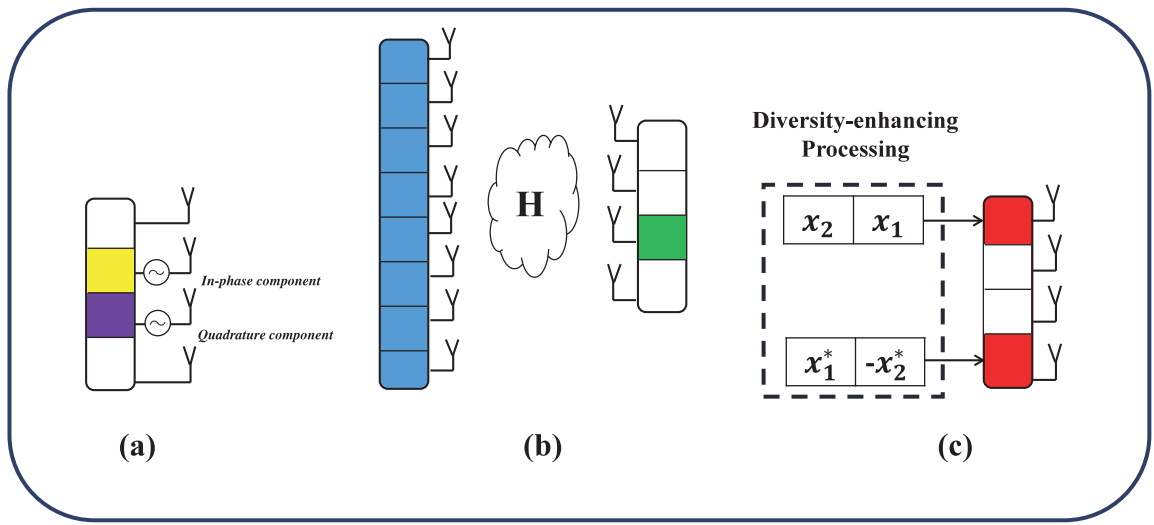}\\
  \caption{Index mapping strategy in enhanced spatial modulation systems of: a) Quadrature Spatial Modulation; b) Precoding-aided Spatial Modulation; c) Space-Time-Block-Coded Spatial Modulation.}\label{sysb}
\end{figure}

Based on the fundamental concept of IM in the space, time, and frequency, various papers  aim at the improvement of the SE/EE, the potential of transmit diversity, precoding and other enhancements of current IM techniques. Due to the similarity among different IM techniques, the enhancements proposed for a specific scheme are usually applicable to all IM techniques. In this section, we introduce some recently proposed  enhancements tailored for IM in order to meet the demanding needs of {5G}. To better illustrate these, in Fig. \ref{sysb}, we present three typical enhancements  designed for SM, which can be readily generalized to other IM schemes.
\subsection{In-Phase/Quadratue IM}
The RF signals  {usually contain} the in-phase component and the quadrature component. The orthogonality between them can be exploited to improve the SE of  IM techniques. For example, given a symbol drawn from the PSK/QAM constellation {\cite{cc8}}, one may map its real part with a cosine carrier into a randomly selected antenna and its imaginary part with a sine carrier to another randomly selected antenna. This technique, termed as quadrature SM { (QSM)}\cite{Mesleh14} (Fig. \ref{sysb} (a)), retains the requirement of a single RF chain while allowing more spatial bits to be transmitted.  The frequency domain counterpart of OFDM with in-phase/quadrature IM (OFDM-IQ-IM), is shown to have significantly better error performance than  IM-OFDM at the same SE \cite{BFMFHY15}.
\subsection{Precoded IM}
IM is not limited to a transmitter design. With {transmitter-side} CSI, effective precoding techniques, such as zero-forcing and MMSE precoders, can  {facilitate} parallel interference-free channels at the receive antennas. Therefore, IM techniques  can be transplanted to the receiver side. By selecting the active receive antenna indices, precoding aided SM (PSM) \cite{cmm} can be formulated as illustrated in Fig. \ref{sysb} (b). Different from SM/G-SM,  the detection of the constellation symbols at different antennas can be performed in a parallel manner, therefore significantly reducing the receiver complexity especially when multiple antennas are  {activated}. Also, in order to bypass channel estimation in PSM, D-SM can also be applied at the receiver side, which also induces 3dB performance penalty.

IM in the space-frequency domain can also be applied at the receiver via precoding. With precoding, the received signals at different receive antennas are free of interchannel interference. As a result, the joint selection of the active elements in space and frequency domains will not increase the detection complexity as at the transmitter side{, which will allow more flexible grouping  across the space and frequency domain to reduce the intra-group channel correlation, and consequently improve the error performance}.
\subsection{Diversity-Enhancing IM}
Beside the SE improvement and the receiver complexity reduction,  research have been dedicated to the error performance enhancement of IM. Among these, diversity-enhancing IM schemes are attracting increasing interests. Since IM are naturally applicable to multiple resources (antennas/subcarriers), some well-known space-time block codes {(STBC)}, such as Alamouti, can be integrated  by carefully designing the mapping of index bits. An example can be found in \cite{c9}. This  STBC-SM scheme is shown in Fig. \ref{sysb} (c).  Such designs are usually migratable among different IM schemes given similar channel properties.  {Specifically,} the coordinate interleaving method proposed for IM-OFDM \cite{c91} is also applicable to all  precoded IM schemes,  because with precoding, the received signal among the receive antennas  {becomes} interference-free.
\subsection{Discussions and Remarks}
{\begin{table*}[!t]
\centering
{
{\caption{Comparison of Existing IM Schemes}}
{\scriptsize
\begin{tabular}{|c|c|c|c|c|c|c|}
         \hline
         Scheme & RF chain number &SE&TX/RX complexity&Diversity order&CSIT/CSIR&Key advantage\\
         \hline
    SM&$1$&Low&Low/Low&$Nr$&No/Yes&High EE \\
    \hline
    D-SM&$1$&Low&Low/Low&$Nr$&No/No& \tabincell{c}{Bypass channel estimation\\and high SE} \\
    \hline
    G-SM&$\leq Nt$&Medium&Medium/Medium&$Nr$&No/Yes&Flexibly SE\\
    \hline
    PSM&$Nt$&Low&High/Low&$Nt-Nr+1$&Yes/Yes&Low Rx complexity\\
    \hline
    QSM&1&Medium&Low/Low&$Nr$&No/Yes&Moderate SE and EE\\
    \hline
    STBC-SM&2&Medium&Medium/Medium&$2Nr$&No/Yes&\tabincell{c}{High reliability\\and moderate EE}\\
    \hline
    V-BLAST(ML)&$Nt$&High&Medium/High&$Nr$&No/Yes&High SE\\
    \hline
    Alamouti&$2$&Medium&Medium/Medium&$2Nr$&No/Yes&High reliability\\
    \hline
    IM-OFDM&$1$&Medium&Low/Low&$Nr$&No/Yes&High EE\\
    \hline
    IM-OFDM-IQ&$1$&Medium&Medium/Medium&$Nr$&No/Yes&High EE and moderate SE\\
    \hline
    MIMO-OFDM-IM&$Nt$&High&High/High&$Nr$&No/Yes&High SE and moderate EE\\
    \hline
    GSFIM&$Nt$&Medium&High/High&$Nr$&No/Yes&Flexible SE \\
    \hline
       \end{tabular}}}
\label{tab1}
\end{table*}
In order to comprehensively compare existing IM schemes, Table I gathers the RF chain number, SE, TX/RX complexity {(computational complexity for encoding and decoding at the TX/RX)}, diversity order, CSI  requirement, and key advantages of the schemes. One can conclude from the table that IM improves the system EE in terms of the hardware cost, computational complexity, and error performance and increases the SE by exploiting the potential of indexing in all available domains.}
\section{Applications of IM}
The implementation of IM in various communication scenarios is also a hot topic. The 5G communication system is expected to accommodate several emerging communication technologies such as massive MIMO, cooperative networks, and full duplex radios as well as supporting the quality-of-service requirements in challenging environments such as the high mobility scenario. In order to better meet such needs, the basic IM techniques should be further tailored and extensive theoretical studies are needed.
\subsection{IM in Massive MIMO}
Massive MIMO has been recognized as one of the key techniques in 5G. With hundreds of antennas at the base station, the SE can be significantly boosted and the multiuser interference can be eliminated. Precoding is a natural way of integrating IM techniques into multi-user massive MIMO downlink \cite{cmm}. Data transmission to multiple users can be accommodated in the same frequency band via space division multiple access, and the detection complexity can also be reduced when multiple antennas are equipped at user terminals, {because the data streams transmitted to the receive antennas are well isolated by linear precoding at the base station}. On the other hand, SM can  be applied at the base station in single user massive MIMO downlink to reduce the number of required RF chains \cite{c10}.
\subsection{IM in Full Duplex Communication}
Full duplex is a promising technique for achieving high SE in wireless systems by transmitting and receiving radio signals simultaneously at the same frequency at a communication node. Consider a communication node equipped only with half-duplex antennas, to facilitate full duplex operation, the node will need multiple antennas and each antenna can not perform reception and transmission simultaneously.  By applying SM at the communication nodes and  using synchronized switches to select the transmit and the receive RF chains to carry additional information, the data rate can be significantly improved without any increase of complexity\cite{YXMLHB13}. This idea is also extended to the decode-and-forward full duplex relay scenarios by selecting the active antenna index at the relay to convey additional information.
\subsection{IM in Cooperative Communication}
Cooperative communication is widely considered to have promising potential of lower cost, improved SE, and extended coverage. Naturally, the combination of IM with cooperative communication is also drawing increasing attention. Generally,  cooperative IM schemes can be classified  according to where IM is applied: 1) dual-hop/multi-hop IM, where IM is applied at the source, and the destination  is served with the help of one or more amplify-and-forward or decode-and-forward relays; 2) distributed IM, where IM is formed at distributed relay nodes according to the information transmitted from the source node; and 3) network coded IM, where IM is utilized at either the source or the relay node in a two-way relay network. Due to the nature of exploiting the antenna index for information transmission, existing studies mainly focus on the potential of applying SM or D-SM in cooperative communication systems \cite{MMXL16}. The combination of cooperative communication with other IM schemes remains to be investigated.
\subsection{IM in High-Mobility Scenarios}
The rapid progress of the high-speed railway systems brings about the urgent need of high data rate wireless communications in high mobility scenarios. Due to the significant Doppler, the accurate estimation of CSI will markedly improve the system complexity and the bandwidth cost. Therefore, noncoherent detection may be more preferable in  high mobility. D-SM is reported to have 3dB performance loss  {than} SM. {With} channel estimation errors and preamble overhead in high mobility scenarios, D-SM may demonstrate improved performance. However, the design of differential IM schemes in other domains still remains to be investigated.

IM schemes  {involving} the frequency domain are also confronted with the challenging problem of ICI  {under} significant Doppler. Careful design of effective ICI cancellation schemes is also an interesting issue. The unique properties of IM, such as the requirement of frequency domain interleaving, render classical ICI cancellation methods not directly applicable. Therefore, talored {ICI cancellation} design  is an interesting problem. Readers may refer to \cite{MXLYXF16} for further information.
\section{Challenges and Future Work}
 As a newly emerging technique, there are abundant appealing and challenging problems to be solved for IM, which we will discuss next.
\subsection{IM in Space}
 IM in space  can also be implemented to groups of antennas instead of individual ones. This applies to cases where the number of transmit antennas is greater than that of receive antennas and should be coupled with precoding, additional information bits can be modulated onto the group indices, therefore enhancing the SE.
\subsection{IM in Frequency}
IM in the frequency domain can also be realized by assigning multiple modes to OFDM subcarriers. In existing IM-OFDM systems, only two modes are permitted, namely null and conventional $M$-ary constellation, and the null mode itself cannot carry any information. Assigning more modes to a subcarrier is an effective means of improving  SE.
\subsection{IM in Time}
IM in the time domain is particularly suitable for single-carrier transmissions and fast fading channels. Recall that the single-carrier frequency domain equalization (SC-FDE) transmitted signal is related to the frequency-domain finite constellation input via IFFT. Discarding some samples of the SC-FDE signal will not induce noticeable effects to the system performance. Since the distortion can be compensated to some extent. The indices of the discarded samples can be then used to carry additional information.
\subsection{Potential Resources for IM}
The existing work on IM inevitably sacrifice some degrees of freedom.  A future direction is the exploration of some alternative resources for IM, for example, the pilot pattern. In high-mobility communications, Pilots occupy a great fraction of space, time, and frequency resources. The exploitation of distinguishable pilot patterns for IM can be therefore an effective means to reduce the SE loss.

Apart from the IM design in contemporary 5G  scenarios, the study of potential applications and  other emerging scenarios such as wireless powered communication systems, simultaneous wireless information and power transfer, and non-orthogonal multiple access also consist of interesting and challenging problems that deserve further investigation. In a multi-access setup, the overall network capacity should be investigated by accounting for the unused resources. This gives rise to another valuable and practical direction for future research.
\section{Conclusions}
In this article, we introduced the concept of IM to improve the SE and EE for next generation communication systems. We have shown that with proper design, IM can be flexibly applied in space, frequency, and time domains. We have discussed various enhancements that are suitable for all IM schemes to further improve the data rate, applicability, and error performance. we also revealed the promising potentials of IM in massive MIMO, high mobility scenarios, and cooperative communication systems. Last but not least, we delineated the opportunities and challenges on the IM design for 5G.
%

\section*{Biographies}
Xiang Cheng (xiangcheng@pku.edu.cn) received the PhD degree from Heriot-Watt University and the University of Edinburgh, Edinburgh, U.K., in 2009, where he received the Postgraduate Research Thesis Prize. He is currently an Associate Professor at Peking University. His general research interests are in areas of channel modeling and communications. Dr. Cheng was the recipient of the IEEE Asia Pacific (AP) Outstanding Young Researcher Award in 2015, and Best Paper Awards at IEEE ITST¡¯12, ICCC¡¯13, and ITSC¡¯14. He has served as Symposium Leading-Chair, Co-Chair, and a Member of the Technical Program Committee for several international conferences. He is now an Associate Editor for IEEE Transactions on Intelligent Transportation Systems.
\\

Meng Zhang (zhangmeng.bri@chinatelecom.cn) received the B.S. degree and the M.S. degree both from the School of Electronics Engineering and Computer Science, Peking University, Beijing, China, in 2014 and 2017, respectively. He is currently an engineer at China Telecom Technology Innovation Center, Beijing, China. His research interests include spatial modulation, OFDM, and wireless powered communications.
\\

Miaowen Wen (eemwwen@scut.edu.cn) received the B.S. degree from Beijing Jiaotong University, Beijing, China, in 2009, and Ph.D. degree from Peking University, Beijing, China, in 2014. From September 2012 to September 2013, he was a Visiting Student Research Collaborator with Princeton University, Princeton, NJ, USA. He is currently an Associate Professor with the South China University of Technology, Guangzhou, China. His research interests include index modulation, underwater acoustic communications, and non-orthogonal multiple access techniques. Dr. Wen was the recipient of Best Paper Awards at IEEE ITST12, ITSC14, and ICNC16. He has served as a Member of the Technical Program Committee for several international conferences. He is now an Associate Editor of the IEEE Access, and on the Editorial Board of the EURASIP journal on Wireless Communications and Networking and the ETRI Journal.
\\

Liuqing Yang (lqyang@engr.colostate.edu) received her Ph.D. degree from the University of Minnesota in 2004. She is currently a Professor at Colorado State University. Her general interests are in areas of communications and signal processing. Dr. Yang was the recipient of the ONR YIP award in 2007, the NSF CAREER award in 2009, the IEEE Globecom Outstanding Service Award in 2010, George T. Abell Outstanding Mid-Career Faculty Award at CSU in 2012, and Best Paper Awards at IEEE ICUWB¡¯06, ICCC¡¯13, ITSC¡¯14, and GLOBECOM¡¯14. She has been actively serving the technical community, including organization of many IEEE international conferences, and the editorial board for a number of journals including IEEE Transactions on Communications, IEEE Transactions on Wireless Communications, and IEEE Transactions on Intelligent Transportation Systems.
\\
\clearpage
\clearpage
\clearpage
\clearpage
\end{document}